\documentclass[letterpaper, 10 pt, conference]{ieeeconf}  

\IEEEoverridecommandlockouts                              

\usepackage{ucs}
\usepackage[utf8x]{inputenc}
\usepackage{amsmath}
\usepackage{amsfonts}
\usepackage{amssymb}
\usepackage[australian]{babel}
\usepackage{fontenc}
\usepackage{graphicx}
\usepackage{yhmath}
\usepackage{epstopdf}
\usepackage{color}
\usepackage{subfigure}
\usepackage{algpseudocode}
\usepackage{algorithm}
\usepackage{comment}
\usepackage{mathrsfs}
\usepackage{booktabs}

\newtheorem{lemma}{Lemma}
\newtheorem{remark}{Remark}

\title{Higher order Voronoi based mobile coverage control}
\author{Bomin Jiang, Zhiyong Sun  and Brian D.O. Anderson 
\thanks{*This work  was supported by the Australian Research Council's Discovery Projects DP-0877562 and DP-110100538 and by NICTA (National ICT Australia. Z. Sun is also supported by the Prime Minister's Australia Asia Incoming Endeavour Postgraduate Award.}
\thanks{ B. Jiang, Z. Sun and B. D. O. Anderson are with Research School of Engineering, The Australian National University. Z. Sun and B.D.O. Anderson are also with Shandong Computer Science Center (National Supercomputer Center in Jinan), Shandong Provincial Key Laboratory of Computer Networks.
        {\tt\small \{u5225976, zhiyong.sun, brian.anderson\}@anu.edu.au}}%
}
\begin{document}
\maketitle

\begin{abstract}
Most current results on coverage control using mobile sensors require that one partitioned cell is the sole responsibility of one sensor. In this paper, we consider a class of generalized Voronoi coverage control problems by using higher order Voronoi partitions, motivated by  applications that more than one senor is required to monitor and cover onecell. We introduce a framework depending on a coverage performance function incorporating higher order Voronoi cells and then design a gradient-based controller which allows the multi-sensor system to achieve a local equilibrium in a distributed manner. In addition, we provide a number of real world scenarios where our framework can be applied. Simulation results are also provided to show the controller performance.
\end{abstract}

\section{Introduction}

In recent years, the performance of distributed multi-agent systems (MASs) in various tasks involving cooperation has been studied with increasing intensity \cite{cao2013overview}. One fundamental problem is that optimal positioning of agents (where \emph{agents} may refer to mobile sensors, vehicles, etc.) to cover an area in a way that some predefined coverage performance function can be optimized. This performance function  can be related to the quality of service of a mobile sensing network, or the cumulative probability of certain events  detected by sensors in the area of interest.
In order to achieve the optimization goal, it is usually required that the control algorithms for different agents are carried out in a decentralized manner \cite{li2005distributed}. That is to say, agents are autonomous, capable of making decisions based on their own information, some of which could be obtained from neighbour agents.
This type of mobile coverage control problem has been studied extensively in the literature since \cite{cortes2004coverage}.
There are many papers on various extensions on this problem including obstacle avoidance \cite{lindhe2005flocking}, non-convex area coverage control \cite{breitenmoser2010voronoi}, coverage control with time-varying density functions \cite{lee2013controlled}, etc.

Typical current mobile coverage control problem is usually solved by using the geometry tool of  Voronoi partitions \cite{de2000computational}. When using this tool, each agent is responsible for monitoring a convex area determined in part by information from its neighbouring agents. Furthermore, each agent moves according to some gradient-based control law to achieve optimization of the coverage performance criterion.

In the framework of coverage control reported in most current literature, each cell in the Voronoi partition is only monitored by one agent.
In this paper we consider a generalization of the classical coverage problem. The starting
idea is to extend `one agent being responsible for one cell' to `two or more agents being responsible for one single cell'. This generalization is motivated by
many real-world applications where, in a coverage task, more than one agents are required to cooperatively monitor one cell. One example is bi-static radar. When deploying the bi-static radar, it is required that the transmitter and receiver are at different locations \cite{4201772}. Though the transmitter is not a sensor, both the transmitter and the receiver need to be reasonably close to a potential target in order to have satisfactory detection; thus the transmitter from the coverage point of view is rather like a sensor.
Another example is the geolocalization problem using TDOA sensors. In that example, at least three sensors at different noncollinear locations are required to localize an object.
The current framework on coverage control is not suitable for the above applications. In this paper, we will use the concept of a higher order Voronoi partition as the main tool. Although this concept is not a recent idea, mobile coverage control using  higher-order Voronoi partitions is novel to the best of our knowledge.

The rest of this paper is organized as follows. Section II reviews the problem settings and current results on the coverage control problem, and further presents some brief introduction to order $k$  Voronoi partition. Section III discusses mainly the order 2 coverage control problem, the controller design,  and the stability analysis of the coverage sensor system. Simulation results are provided in Section IV to show the coverage properties of the controller. In Section V, we shows some potential applications of the high order coverage framework.  Finally, Section VI concludes this paper.

\section{Background literature}
\subsection{Order one Voronoi partition and coverage control}
\label{Background}
Suppose there is a 2-D convex area $Q$ to be covered by $n$ mobile sensors in this area. Note the convexity assumption we make in this paper is common in many research papers on coverage control, e.g. \cite{cortes2004coverage, schwager2009decentralized}; there are also some papers which focus on non-convex region coverage control (see e.g. \cite{breitenmoser2010voronoi}). A point in $Q$ is denoted as $q$ and sensor $i$'s position is denoted by $p_i \in \mathbb{R}^2$. A coverage performance function $\mathcal{H}(p_1, p_2,\cdots,p_n)$ is defined as follows
\begin{equation}
\label{single performance define}
\mathcal{H}(p_1, p_2,\cdots,p_n)=\int_Q \min_{i\in \{1,\cdots,n\}} f(\|q,p_i\|) \phi(q)dq
\end{equation}
where $\phi$ is a distribution density function known to all sensors, $\|q,p_i\|$ denotes the Euclidean distance between $q$ and $p_i$, and
the function $f(\|q,p_i\|)$ describes the measurement cost or measurement quality at a point $q$ by a sensor at $p_i$. We also suppose that $f$ should be monotonically increasing and differentiable. The coverage control aims to  minimize the above performance function and to find the corresponding optimal positions of mobile agents.

The minimum inside the integral of the performance function \eqref{single performance define} induces a partition of $Q$ into non-overlapping cells. These cells   are called the Voronoi partition $\{V_1,\cdots,V_n\}$ of $Q$ generated by the points $p_1, p_2,\cdots,p_n$ defined as
\begin{equation}
V_i=\{q\in Q | ~\| q,p_i\| \leq \| q,p_j\|,\forall j\neq i\}
\end{equation}
For a given set of $p_1, p_2,\cdots,p_n$, if $V_i$ and $V_j$ are adjacent (i.e. two cells which have a boundary either comprising an interval of nonzero length or which may just be a single point; Whichever is appropriate should be used), then agents  $i$ and  $j$ are defined as neighbours. The neighbor set of agent $i$ is denoted as $\mathcal{N}_i$.
By using the Voronoi partition, the performance function can be transformed as
\begin{equation}
\int_Q \min_{i\in \{1,\cdots,n\}} f(\|q,p_i\|) \phi(q)dq=\sum_{i=1}^n \int_{V_i} f(\|q,p_i\|) \phi(q)dq
\end{equation}
The controller proposed in e.g. \cite{cortes2004coverage} is a gradient-based controller minimizing the performance function. An optimal coverage performance can be obtained by moving mobile sensor positions in accordance with the gradient-based law. The optimum may be local, not global.

\subsection{Order $k$ Voronoi partition}
Most literature on coverage control, such as the works reviewed in Section II.A, assumes that each sensor is responsible for sensing or monitoring   its own region.
We term this an \emph{order one coverage control} problem.
Now we are going to generalize the problem to a higher order coverage problem such that each cell is defined by two or more sensors.
The general literature of order $k$ Voronoi partitions includes \cite{aurenhammer1991voronoi,boissonnat1993semidynamic}. Note there are mature algorithms to compute the order $k$ Voronoi partition of a given area; see the survey paper \cite{aurenhammer1991voronoi} or \cite{agarwal1998constructing}.

The definition of an order $k$ Voronoi partition of a convex area $Q$ is given below following \cite{agarwal1998constructing}. Let $\mathcal{S}$ be the set of sensors' positions in $Q$. Suppose further that $\mathcal{T}$ is a subset of $\mathcal{S}$ and there are $k$ elements in $\mathcal{T}$.  The generalized Voronoi partition is defined as
\begin{equation}\label{define order k}
V(\mathcal{T})=\{q |\forall v\in \mathcal{T}, \forall w\in \mathcal{S} \backslash \mathcal{T}, \|(q,v)\| \leq \|(q,w)\|,|\mathcal{T}| = k\}
\end{equation}
where $\mathcal{S}\backslash\mathcal{T}$ denotes the relative complement of $\mathcal{T}$ with respect to $\mathcal{S}$. For each point $q$ in $V(\mathcal{T})$, $q$ is not further to any sensor in $\mathcal{T}$ than to any sensor not in $\mathcal{T}$.

\begin{figure}[!ht]
\begin{center}
\includegraphics[width=0.7\columnwidth]{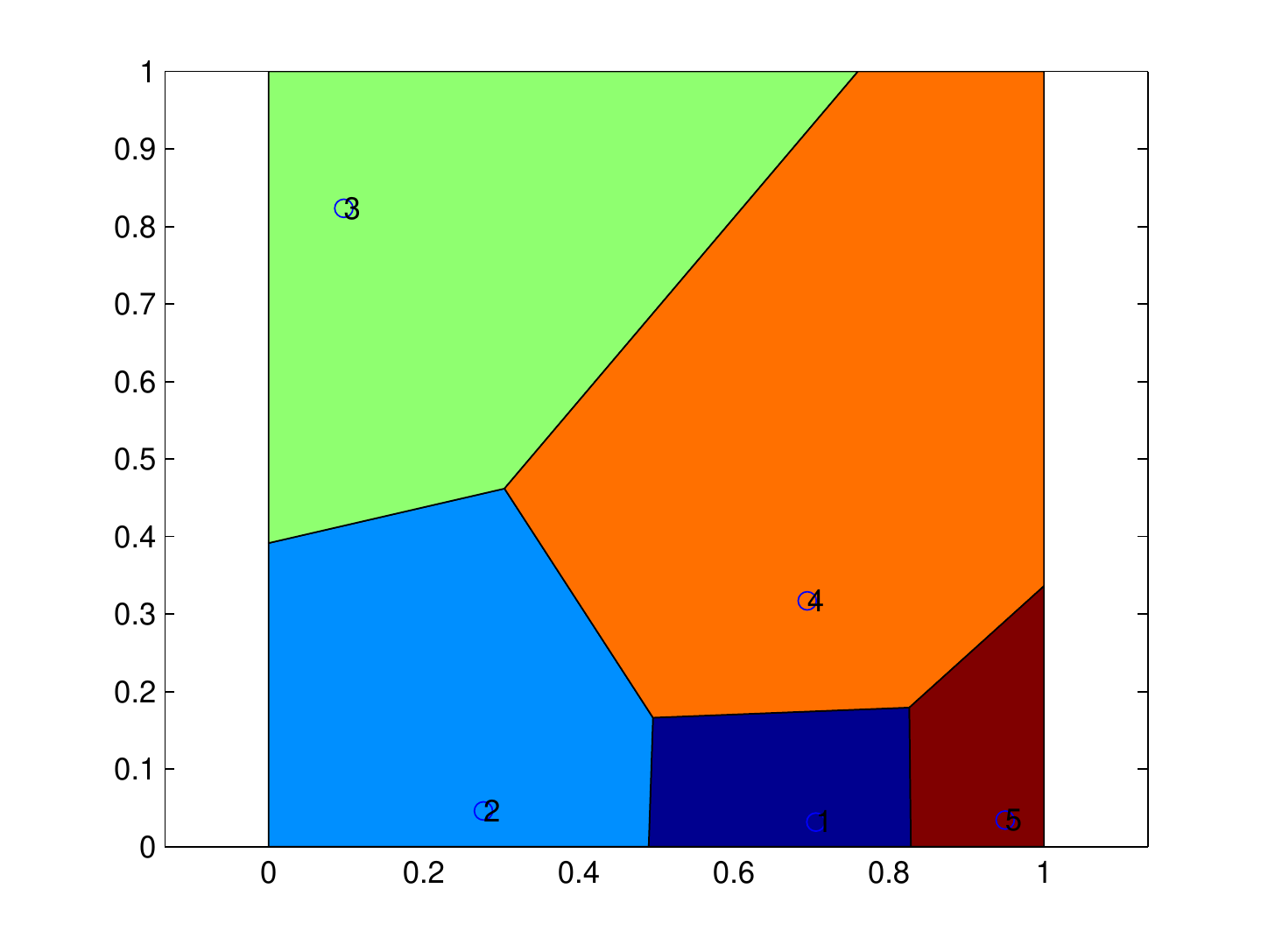}
\end{center}
\caption{An example of an order 1 Voronoi partition}
\label{fig1}
\end{figure}

\begin{figure}[!ht]
\begin{center}
\includegraphics[width=0.7\columnwidth]{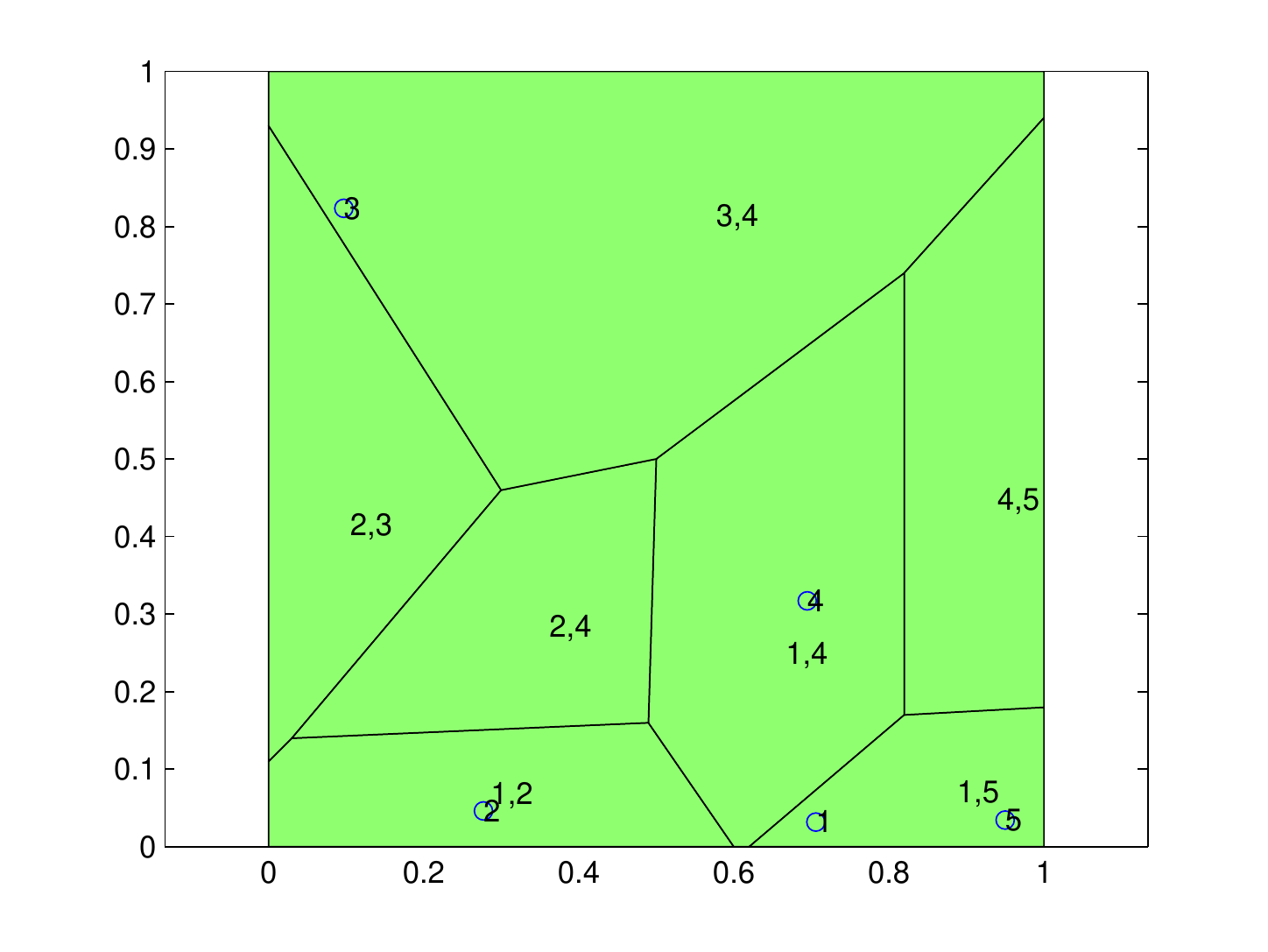}
\end{center}
\caption{An example of an order 2 Voronoi partition with the same sensor positions with Figure \ref{fig1}}
\label{fig2}
\end{figure}

In this paper we largely focus on the order 2 coverage problem. 
As an example, Figure \ref{fig1} shows an order 1 Voronoi partition and Figure \ref{fig2} shows a corresponding order 2 Voronoi partition with the same sensor positions as Figure \ref{fig1}. There are some similar properties in higher-order partitions in comparison to order 1 Voronoi partitions, and we list some: (i) no two cells of $V(\mathcal{T})$ overlap; (ii) the union of the cells, which are all closed, covers the convex region $Q$, and (iii) not every $k$-combination of sensors necessarily defines a cell in the partition.

For all $p_i\in \mathcal{S}$, there are some $\mathcal{T}$ that contain $p_i$. We put all these $\mathcal{T}$ into a set $\mathcal{P}_i$ so that $\mathcal{P}_i=\{\mathcal{T}| \mathcal{T} \subset \mathcal{S},p_i\in\mathcal{T}\}$. Further suppose that $W_i=\cup_{\mathcal{T}\in\mathcal{P}_i} V_{\mathcal{T}}$. It is noticeable that when we put these $\mathcal{T}$ together, we will not obtain the same cell containing $p_i$ in the order 1 partition. In fact, there holds $p_i\in V_i\subset W_i$. In addition, according to the definition of higher order Voronoi partition, $V_i$ and $V_{\mathcal{T}}$ are both always convex but $W_i$ may not be convex. Other properties of higher order Voronoi partitions can be found in \cite{chazelle1987improved} and \cite{lee1982k}. 


\section{Order two coverage control}
\label{order two section}
\subsection{Performance function and its relationship with generalized Voronoi partition}
In this section, we are going to discuss order two coverage control so throughout this section we assume that $k=2$. Define a set $C=\{i,j|~i,j\in\{1,\cdots,n\},i < j\}$. We also note that the set designation of a particular $\mathcal{T}$ as $\mathcal{T}_{ij}$ means points $p_i,p_j \in \mathcal{T}_{ij}$ and $(i,j)\in C$.
The generalized sensing performance function $\mathcal{H}(p_1, p_2,\cdots,p_n)$ is constructed  as follows
\begin{equation}
\label{double cover}
\mathcal{H}(p_1, p_2,\cdots,p_n)=\int_Q \min_{(i,j)\in C} f(\|q,p_i\|,\|q,p_j\|) \phi(q) dq
\end{equation}

We are trying to find optimal positions of the sensors that can minimize the above performance function.
Similarly to an order 1 Voronoi coverage problem, the function $f(\cdot, \cdot)$  indicates the measurement quality of a point $q$ but now by a pair of agents.
Therefore, for a set of fixed sensor positions, we might measure the quality of sensing associated with those positions by $\int_Q \min_{(i,j)\in C} (\|q,p_i\|+\|q,p_j\|) \phi(q) dq$.  In fact, there is a broad set of $f(\cdot, \cdot)$ for which it makes sense to formulate such a measure. We shall in fact impose certain properties on $f(\cdot, \cdot)$
analogous to the order 1 mobile coverage control case, where $f(\cdot)$ being monotonically increasing and differentiable is a basic requirement. In the order 2 case, besides requiring that the function $f(\cdot, \cdot)$ should be differentiable, it should also have the following properties

\begin{enumerate}
  \item $\frac{\partial}{\partial \|q,p_i\|}f(\|q,p_i\|,\|q,p_j\|)\geq 0$ \\
  \item $\frac{\partial}{\partial \|q,p_j\|}f(\|q,p_i\|,\|q,p_j\|)\geq 0$, and \\
  \item $f(\|q,p_i\|,\|q,p_j\|)=f(\|q,p_j\|,\|q,p_i\|)$
\end{enumerate}
The first two properties correspond to the monotonically increasing property in the order 1 case. We note that the higher order Voronoi partition is defined as
\eqref{define order k}, where the order of the elements in $\mathcal{T}$ should not affect the actual partition. As a result,  the order of independent variables should not affect the value of $f(\cdot, \cdot)$, either. The third requirement in the above condition ensures this property.

Based on the above performance function \eqref{double cover} and the distance function, we can obtain the following two lemmas.

\begin{lemma}
\label{lemma1} With the definitions of $Q$, $C$, $f(\cdot, \cdot)$ $\mathcal{T}_{ij}$, $\mathcal{S}$ and $V_{\mathcal{T}_{ij}}$ showed above,
for all $q$ in the set $V_{\mathcal{T}_{ij}}$ and  $(k,l)\neq(i,j)$, there holds \begin{equation} f(\|q,p_i\|,\|q,p_j\|) \leq f(\|q,p_k\|,\|q,p_l\|)\end{equation}
\end{lemma}

\parskip= 6pt
\begin{proof}
According to the definition of $V_{\mathcal{T}_{ij}}$, we know each $\|q,p_i\|$ and $\|q,p_j\|$ is less than or equal to both $\|q,p_k\|$ and $\|q,p_l\|$. Because $\frac{\partial}{\partial \|q,p_i\|}f(\|q,p_i\|,\|q,p_j\|)\geq 0$, there holds $f(\|q,p_i\|,\|q,p_j\|)\leq f(\|q,p_k\|,\|q,p_j\|)$.

Further because $\frac{\partial}{\partial \|q,p_j\|}f(\|q,p_k\|,\|q,p_j\|)\geq 0$, there also holds $f(\|q,p_k\|,\|q,p_j\|)\leq f(\|q,p_k\|,\|q,p_l\|)$. Thus the lemma is proved.
\end{proof}

\begin{lemma} By using the higher order Voronoi partition and the distance function defined above, the performance function can be further transformed as
\begin{equation}
\label{double min achieve}
\begin{split}
\mathcal{H}=&\int_Q \min_{(i,j)\in C}f(\|q,p_i\|,\|q,p_j\|) \phi(q) dq\\
=&\sum_{\forall\mathcal{T}_{ij}\subset\mathcal{S}}\int_{V_{\mathcal{T}_{ij}}}f(\|q,p_i\|,\|q,p_j\|) \phi(q) dq\\
\end{split}
\end{equation}
\end{lemma}
This lemma is a straightforward consequence of Lemma \ref{lemma1}.

According to Lemma \ref{lemma1}, as long as the three properties of $f(\cdot,\cdot)$ hold, one can transform the original performance function \eqref{double cover} into \eqref{double min achieve} by using the order-2 Voronoi concept.
Different $f(\cdot,\cdot)$ in the performance function will affect directly the optimization of the performance function, and thus the final configuration of sensors' positions.
The choice of $f(\cdot,\cdot)$ depends on specific requirements for different applications. Here are some typical $f(\cdot,\cdot)$, expressed in terms of norm.
\begin{enumerate}
  \item When $f(\cdot,\cdot)=\|q,p_i\|+\|q,p_j\|$, the sensing performance for each $q\in V_{\mathcal{T}_{ij}}$ is related to the distance sum from $q$ to two sites $p_i$ and $p_j$.  The detailed explanation of these applications will be discussed later in Section \ref{Real world applications}.
  \item When $f(\cdot,\cdot)=\|q,p_i\|^2+\|q,p_j\|^2$, the sensing performance is expressed by the squared distance sum to two sites $p_i$ and $p_j$.  In this case, as we are going to show in the next section, the controller expression has a strong relationship with the centroid of each cell.
  \item Suppose $f(\cdot,\cdot)=(\|q,p_i\|^n+\|q,p_j\|^n)^{1/n}$. As $n\rightarrow \infty$, there holds $f(\cdot,\cdot)=\max\{\|q,p_i\|,\|q,p_j\|\}$.
\end{enumerate}
The three examples just listed correspond to the order 2 case but the idea can be generalized to higher order cases.
For the generalized performance function \eqref{double cover}, the optimal positions of sensors are $f$ dependent. In Section V, we will provide more discussion on specific applications by using different  $f(\cdot,\cdot)$.

\subsection{Controller design}

\subsubsection{Gradient-based controller}
\label{Gradient-based controller}
In order to minimize the performance function \eqref{double min achieve}, we can design a controller for each sensor with position $p_i$ as
$$ \dot p_i=-\frac{\partial \mathcal{H}}{\partial {p_i}}$$
The mobile sensor system with the above controller defines a gradient flow of the performance function \eqref{double cover}. According to the property of  gradient systems \cite{absil2006stable}, the above gradient controller provides a natural choice to optimize the performance function.
Furthermore, in order to implement the above controller, each agent needs to know the position of its neighbouring agents.
The neighbouring agents of $i$ are $\mathcal{N}_i = \{j|\mathcal{T}_{ij}\in \mathcal{P}\}\cup\{k,l|V_{\mathcal{T}_{ij}}\cap V_{\mathcal{T}_{kl}} \neq \emptyset \}$, which are agents that monitor the cells $V_{\mathcal{T}_{ij}}$ and the cells with common boundaries with these $V_{\mathcal{T}_{ij}}$.

\subsubsection{Cancellation of boundary terms}
In the following, we will present an explicit formula for the controller. In our order two problem, for each $p_i\in\mathcal{S}$, we are going to show
\begin{equation}\label{pupp}
\frac{\partial \mathcal{H}}{\partial {p_i}}=\sum_{\forall \mathcal{T}_{ij},p_i\in\mathcal{T}_{ij}}
\int_{V_{\mathcal{T}_{ij}}}\frac{\partial}{\partial {p_i}}f(\|q,p_i\|,\|q,p_j\|) \phi(q) dq
\end{equation}

In the expression of $\mathcal{H}$ as shown in \eqref{double min achieve}, the domain of integration is a function of $p_i$. As a result, when one calculates the partial derivative of $\mathcal{H}$ with respect to $p_i$, one needs to deal with the problem of differentiation under the integral sign. Some basic facts about the problem of differentiating under the integral sign are given in the Appendix.

\begin{figure}[!ht]
\begin{center}
\includegraphics[width=0.7\columnwidth]{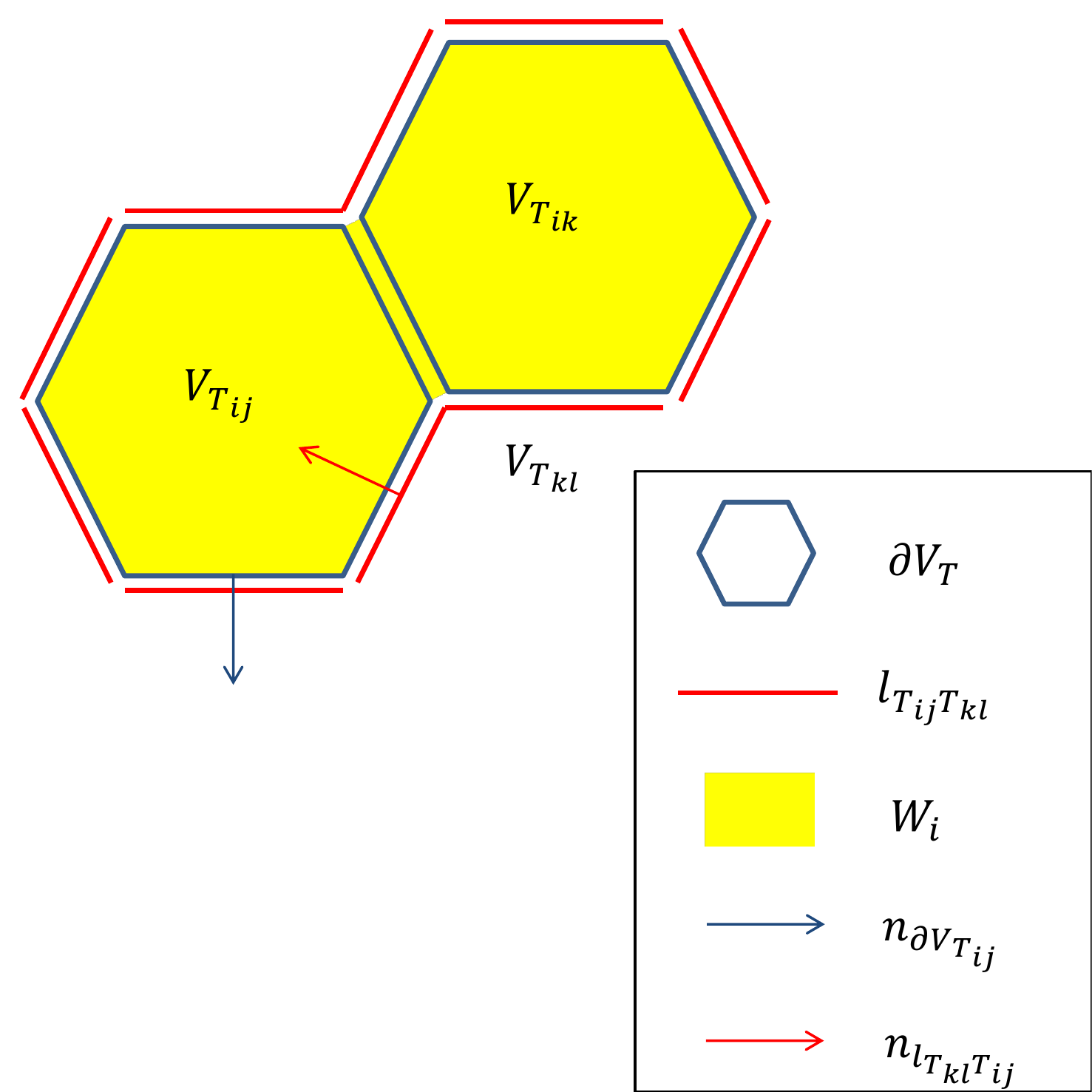}
\end{center}
\caption{Graphical representations of notations referred to each cell and the boundary}
\label{fig_notation}
\end{figure}

Now we are going to define some quantities in relation to order 2 Voronoi cells. Figure \ref{fig_notation} shows  illustrative representations of the notations used in the derivation below.
Let $\partial V_{\mathcal{T}_{ij}}$ and $\partial Q$ be the boundary of $V_{\mathcal{T}_{ij}}$ and $Q$ respectively. Further suppose that $l_{\mathcal{T}_{ij}\mathcal{T}_{kl}}=V_{\mathcal{T}_{ij}}\cap V_{\mathcal{T}_{kl}}$ is the common boundary of $V_{\mathcal{T}_{ij}}$ and $V_{\mathcal{T}_{kl}}$. Let $q_{\partial V_{\mathcal{T}_{ij}}}$ be a point on the boundary of $V_{\mathcal{T}_{ij}}$, and $q_{l_{\mathcal{T}_{ij}\mathcal{T}_{kl}}}$ be a point on the common boundary of $V_{\mathcal{T}_{ij}}$ and $V_{\mathcal{T}_{kl}}$. We also define $n_{\partial V_{\mathcal{T}_{ij}}}$ as the outward facing unit normal vector of $\partial V_{\mathcal{T}_{ij}}$, and $n_{l_{\mathcal{T}_{ij}\mathcal{T}_{kl}}}$ as the unit normal vector of $l_{\mathcal{T}_{ij}\mathcal{T}_{kl}}$ from cell $\mathcal{T}_{ij}$ to $\mathcal{T}_{kl}$.

In our problem, if we calculate the partial derivatives of $\mathcal{H}$ with respect to each sensor position $p_i$ under the integral sign, we have

\begin{equation}\label{puppder}
\begin{split}
\frac{\partial \mathcal{H}}{\partial {p_i}}&=\sum_{\mathcal{T}_{ij}\in \mathcal{P}_i}
\int_{V_{\mathcal{T}_{ij}}}\frac{\partial}{\partial {p_i}}f(\cdot,\cdot) \phi(q) dq\\
&+\sum_{\mathcal{T}_{ij}\in \mathcal{P}_i}
\int_{\partial Q \cap \partial V_{\mathcal{T}_{ij}}} f(\cdot,\cdot) \phi(q) \frac{\partial q_{ \partial V_{\mathcal{T}_{ij}}}}{\partial p_i} n_{\partial V_{\mathcal{T}_{ij}}} dq\\
&+\sum_{\mathcal{T}_{ij}\in \mathcal{P}_i}
\int_{\partial V_{\mathcal{T}_{ij}} \backslash \partial Q} f(\cdot,\cdot) \phi(q) \frac{\partial q_{\partial V_{\mathcal{T}_{ij}}}}{\partial p_i} n_{\partial V_{\mathcal{T}_{ij}}} dq\\
&+\sum_{\underset{V_{\mathcal{T}_{ij}}\cap V_{\mathcal{T}_{kl}}\neq \emptyset}{\mathcal{T}_{ij}\in \mathcal{P}_i,\mathcal{T}_{kl}\not\in \mathcal{P}_i}}
\int_{l_{\mathcal{T}_{ij}\mathcal{T}_{kl}}} f(\cdot,\cdot) \phi(q) \frac{\partial q_{l_{\mathcal{T}_{ij}\mathcal{T}_{kl}}}}{\partial p_i} n_{l_{\mathcal{T}_{kl}\mathcal{T}_{ij}}} dq\\
\end{split}
\end{equation}

Note the second line in \eqref{puppder} is always zero because $\partial Q$ is always stationary and its partial derivatives with respect to the entries of $p_i$ are always zero. By the definition of higher order Voronoi cell \cite{shamos1975closest}, we know $l_{\mathcal{T}_{ij}\mathcal{T}_{kl}}=l_{\mathcal{T}_{kl}\mathcal{T}_{ij}}$ and $n_{l_{\mathcal{T}_{ij}\mathcal{T}_{kl}}}=-n_{l_{\mathcal{T}_{kl}\mathcal{T}_{ij}}}$. As a result, the third and fourth lines in \eqref{puppder} cancel out with each other. Therefore \eqref{pupp} holds.

\subsubsection{Relationship with cell centroids}

For any given $f(\|q,p_i\|,\|q,p_j\|)$, it is not easy to find the general relationship of \eqref{pupp} with cell centroids of $W_i$, so we first consider the case when
$f(\|q,p_i\|,\|q,p_j\|)=\frac{1}{2}(\|q,p_i\|^2 + \|q,p_j\|^2)$. In this case, the performance function becomes $$\mathcal{H}(p_1, p_2,\cdots,p_n)=\int_Q \min_{(i,j)\in C} \frac{1}{2}(\|q,p_i\|^2 + \|q,p_j\|^2) \phi(q) dq$$

Note the centroid and mass of a Voronoi cell $V_{\mathcal{T}_{ij}}$ are
$$C_{V_{\mathcal{T}_{ij}}}=\frac{\int_{V_{\mathcal{T}_{ij}}}q\phi(q)dq}{\int_{V_{\mathcal{T}_{ij}}}\phi(q)dq}$$
and $${M}_{V_{\mathcal{T}_{ij}}}=\int_{V_{\mathcal{T}_{ij}}}\phi(q)dq$$ respectively.
In this case,
\begin{equation}\label{puppcentroid}
\begin{split}
\frac{\partial \mathcal{H}}{\partial {p_i}}&=\sum_{\mathcal{T}_{ij}\in \mathcal{P}_i}
\int_{V_{\mathcal{T}_{ij}}}\frac{\partial}{\partial {p_i}}\frac{1}{2}(\|q,p_i\|^2 + \|q,p_j\|^2) \phi(q) dq\\
&=\sum_{\mathcal{T}_{ij}\in \mathcal{P}_i}\int_{V_{\mathcal{T}_{ij}}}(q-p_i)\phi(q)dq\\
&=\sum_{\mathcal{T}_{ij}\in \mathcal{P}_i}-{M}_{V_{\mathcal{T}_{ij}}}(C_{V_{\mathcal{T}_{ij}}}-p_i)\\
\end{split}
\end{equation}

Suppose further that the centroid and mass of $W_i$ are
$$C_{W_i}=\frac{\sum_{\mathcal{T}_{ij}\in \mathcal{P}_i}{M}_{V_{\mathcal{T}_{ij}}}C_{V_{\mathcal{T}_{ij}}}}{\sum_{\mathcal{T}_{ij}\in \mathcal{P}_i}{M}_{V_{\mathcal{T}_{ij}}}}$$
and $${M}_{W_i}=\sum_{\mathcal{T}_{ij}\in \mathcal{P}_i}{M}_{V_{\mathcal{T}_{ij}}}$$ respectively.
Now we have
\begin{equation}\label{puppcentroidW}
\begin{split}
\frac{\partial \mathcal{H}}{\partial {p_i}}&=\sum_{\mathcal{T}_{ij}\in \mathcal{P}_i}-{M}_{V_{\mathcal{T}_{ij}}}(C_{V_{\mathcal{T}_{ij}}}-p_i)\\
&=-C_{W_i}\sum_{\mathcal{T}_{ij}\in \mathcal{P}_i}{M}_{V_{\mathcal{T}_{ij}}}+p_i\sum_{\mathcal{T}_{ij}\in \mathcal{P}_i}{M}_{V_{\mathcal{T}_{ij}}}\\
&=-{M}_{W_i}(C_{W_i}-p_i)\\
\end{split}
\end{equation}
The above result indicates that $p_i$ is moving towards the centroid of $W_i$.
Note that this centroid in general moves when the sensors move, since the Voronoi diagram will change with moving sensors.
This result is very similar to the order 1 centroid coverage control case when $f(x)$ is defined as $f(x) = \frac{1}{2}x^2$. 
Now we summarize the convergence results concerning the above gradient system as follows.

\begin{lemma}
For a group of mobile agents with the closed-loop system induced by  \eqref{pupp}, all the agents' positions  will  converge to
the set of critical points of $\mathcal{H}$. Furthermore, by taking $f(\|q,p_i\|,\|q,p_j\|)=\frac{1}{2}(\|q,p_i\|^2 + \|q,p_j\|^2)$ and designing the controller as $\dot p_i =-K(C_{W_i}-p_i)$ where $K$ is a positive gain, agents' locations will converge to the cell centroids which result in a higher order  centroidal  Voronoi configuration.
\end{lemma}

The proof follows directly from the properties of gradient systems (see e.g. \cite{absil2006stable}) and is omitted here.

\begin{remark}
In this section we mainly focus on generalized coverage control using an order 2 Voronoi partition. We mention that the coverage scheme can be extended to higher order Voronoi partitions by modifying the performance function stated in \eqref{double cover} with reasonable distance functions. Also, the main analysis for the 2-D case can be extended straightforwardly to 3-D space coverage. The controller for the 3-D case takes the same form as in \eqref{pupp} but the derivation requires a more sophisticated analysis of boundary issues.
\end{remark}
\begin{remark}
The discretized version of the original order 1 coverage control problem in the field of data analysis and image processing is called k-mean clustering \cite{du1999centroidal}. To the best of our knowledge, there is no previous literature on the k-mean clustering problem corresponding to the `order 2' coverage problem in the continuous case. In the future, we intend to use the idea of higher order Voronoi partition in the research of k-mean clustering methods.
\end{remark}

\section{Simulation results and discussion}
%
%
%

\begin{figure}[!ht]
\subfigure[Initial sensor positions and order 2 Voronoi partition]{
\includegraphics[width=0.7\columnwidth]{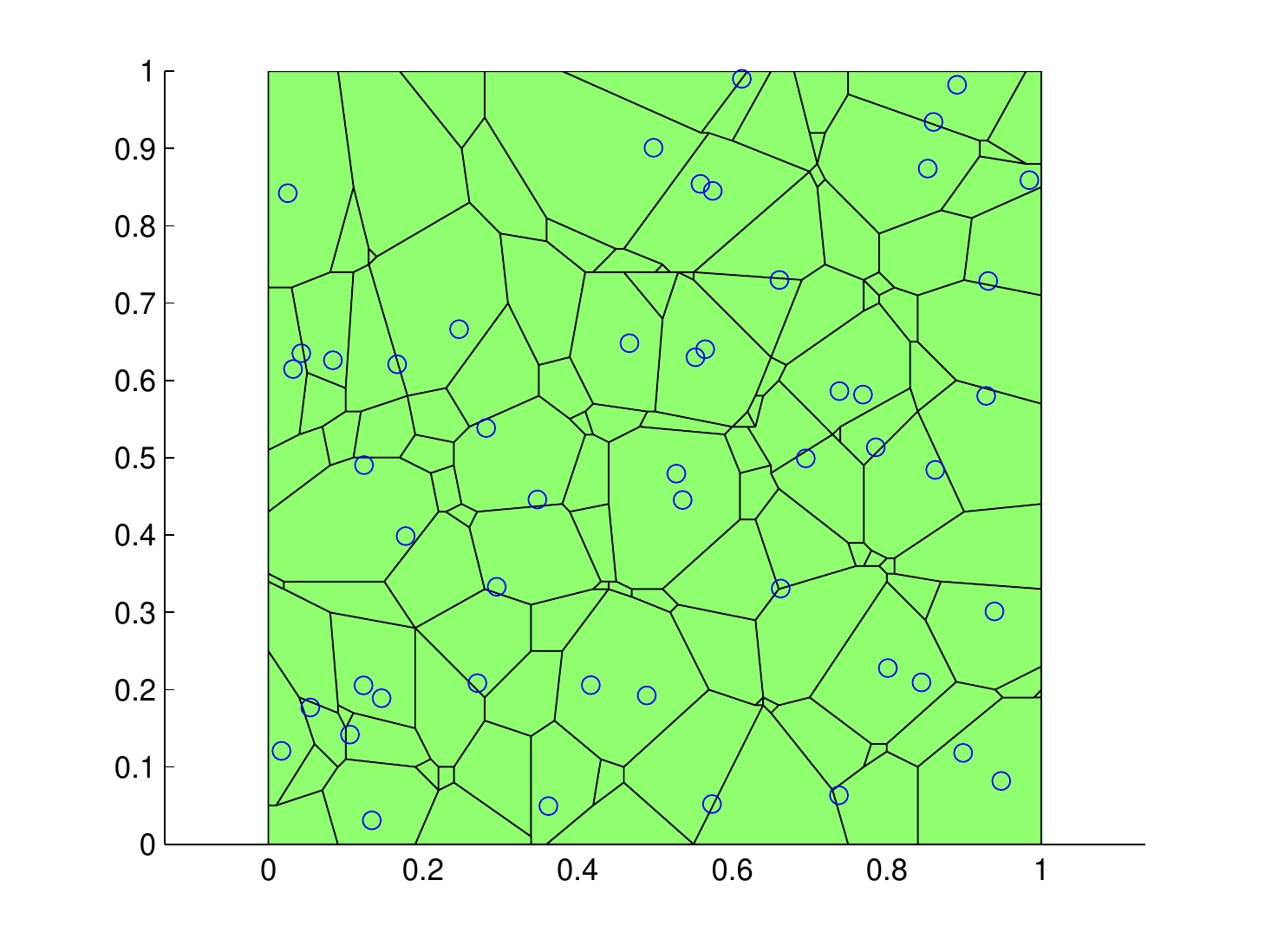}
\label{s11}
}
\subfigure[Moving trajectory of sensors]{
\includegraphics[width=0.7\columnwidth]{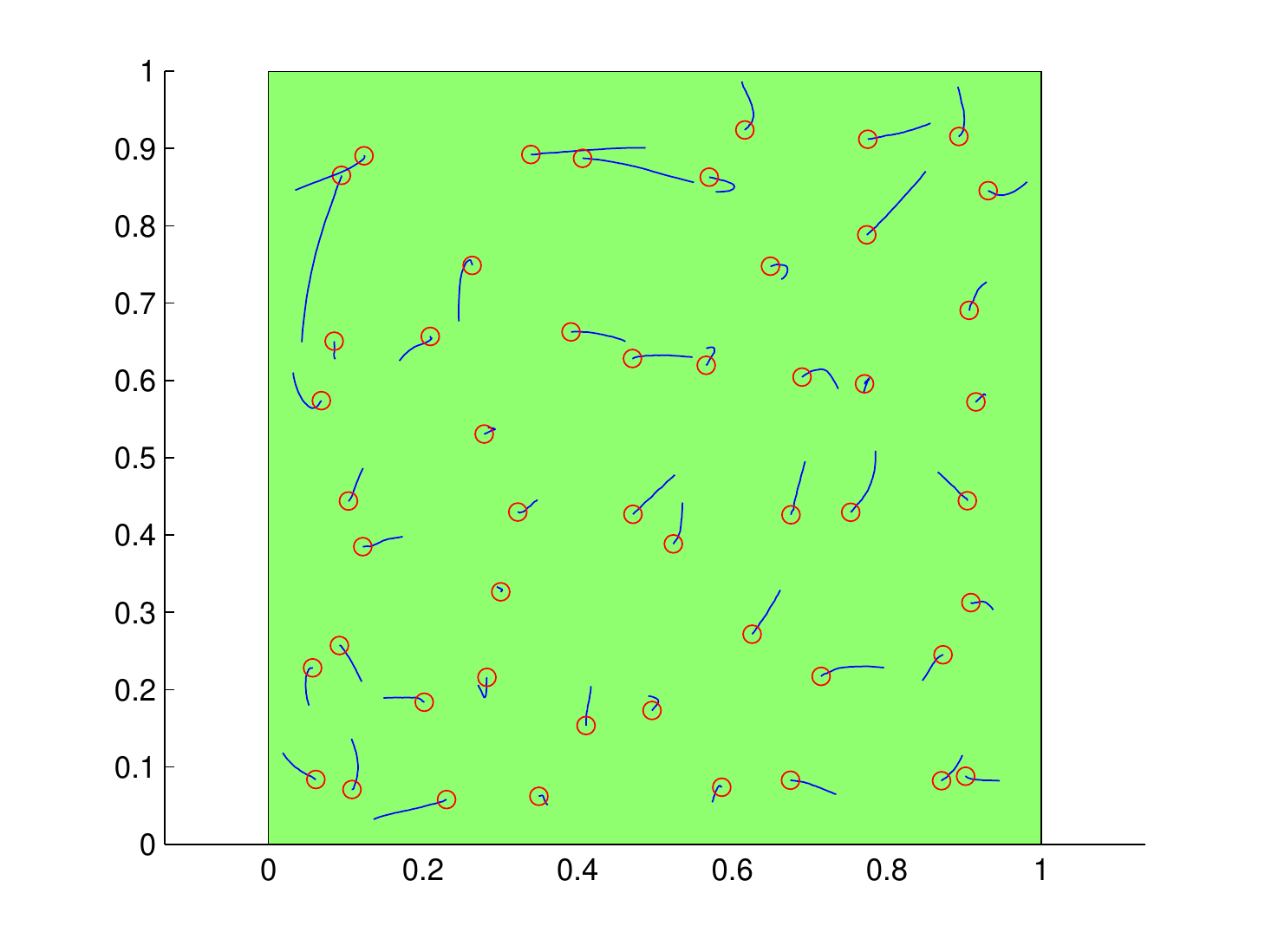}
\label{s12}
}
\subfigure[Final sensor positions and order 2 Voronoi partition]{
\includegraphics[width=0.7\columnwidth]{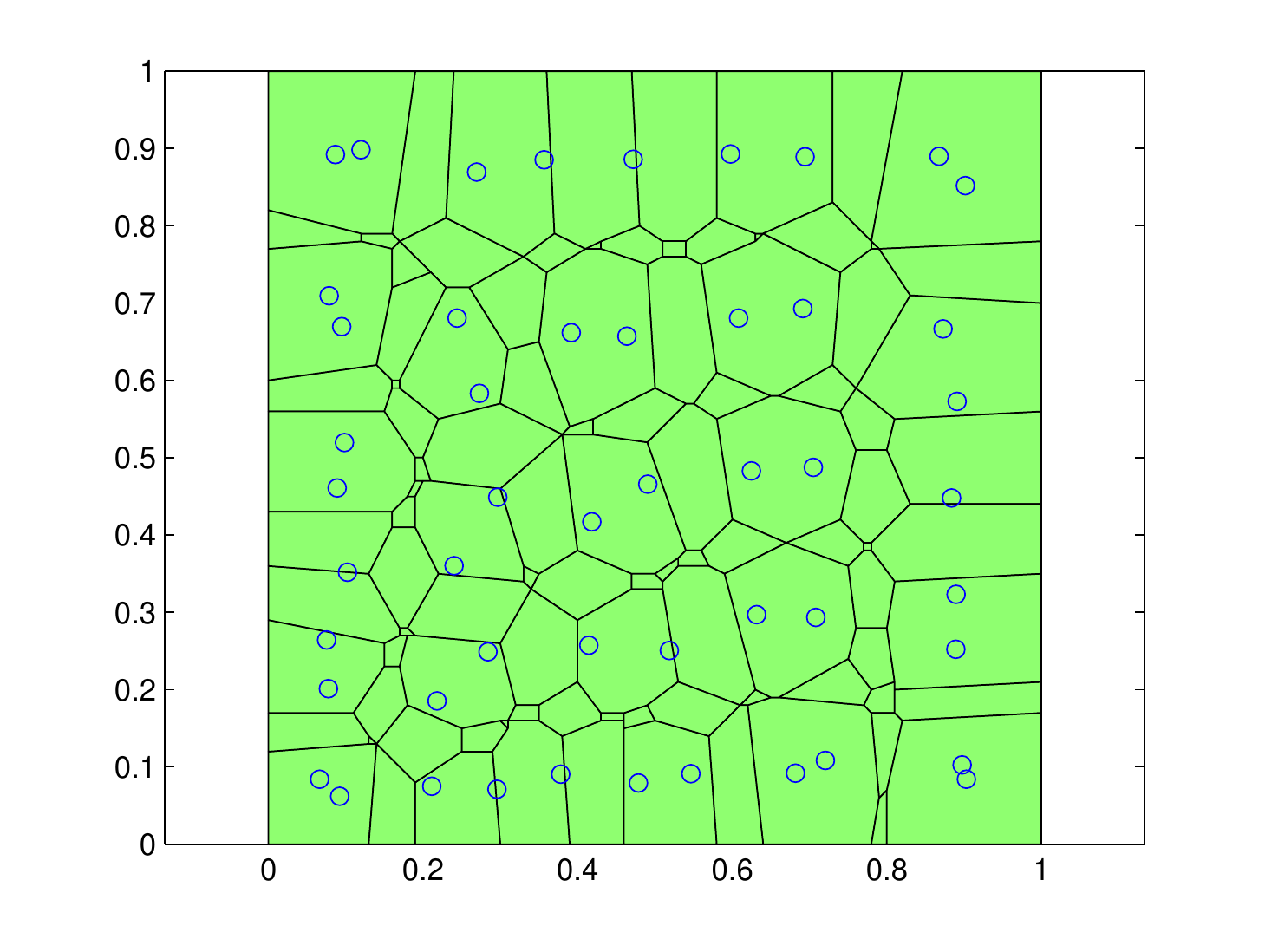}
\label{s13}
}
\subfigure[Evolution of the coverage objective function]{
\includegraphics[width=0.7\columnwidth]{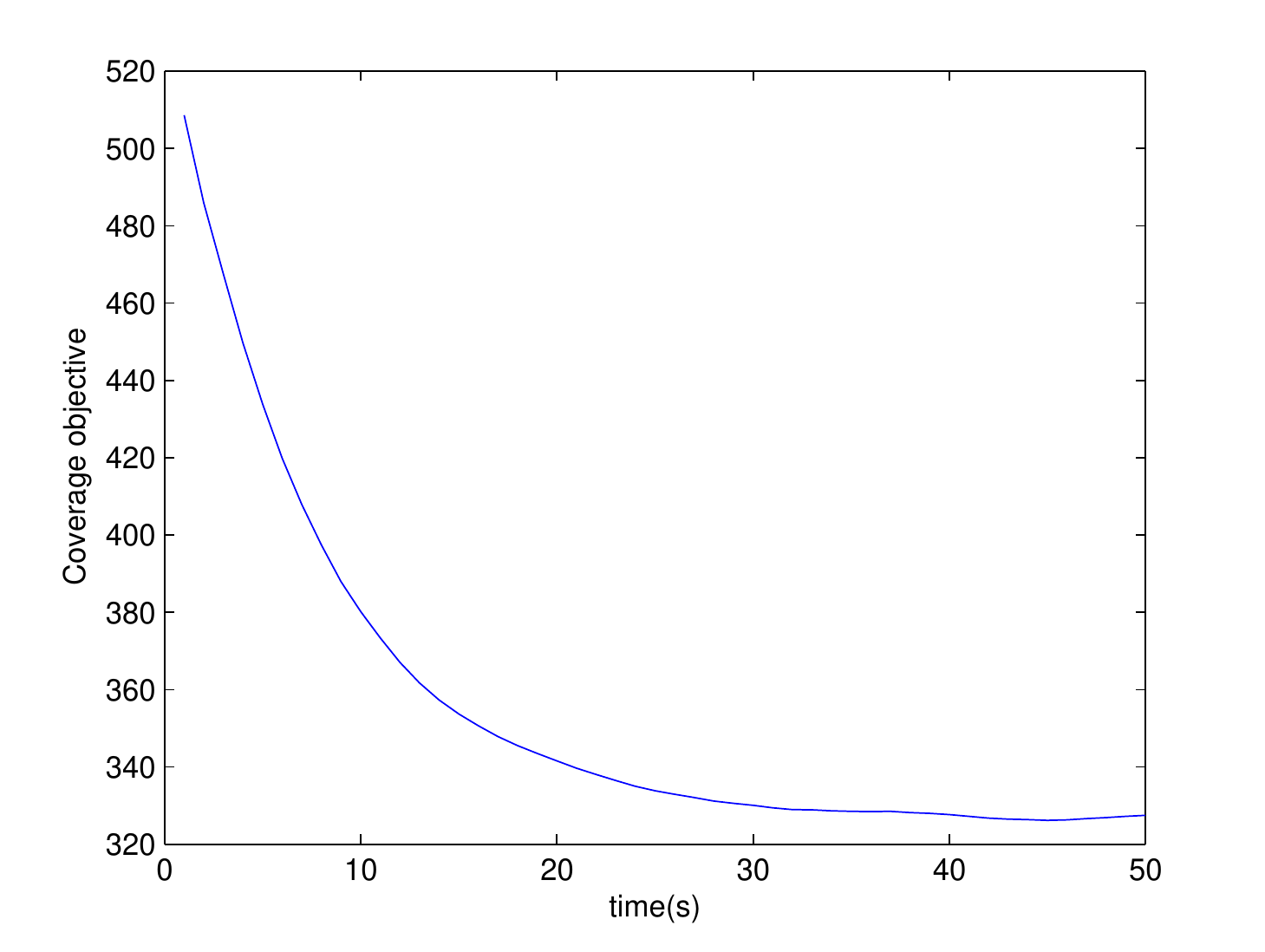}
\label{s14}
}
\caption{A simulation result with 50 mobile sensors}\label{50SensorResult}
\end{figure}

A simulation with 50 agents is shown in Figure \ref{50SensorResult}. Figure \ref{s11} shows the initial positions of a group of agents. The initial positions of these agents are generated randomly with a uniform distribution in the green square area. This figure also shows the order 2 Voronoi partition with these initial positions. Apart from that, Figure \ref{s12} shows the moving trajectory of these agents using the controller designed in Section \ref{Gradient-based controller}. Note that $f(\cdot,\cdot)=\frac{1}{2}\|q,p_i\|^2+\|q,p_j\|^2$ and $\phi(q)=1$. Each blue curve represents the trajectory of one agent and each red circle represents the final position of the agent. In addition, Figure \ref{s13} shows the final positions of the group of agents and also the final order 2 Voronoi partitions. Furthermore, Figure \ref{s14} shows the evolution of the value of the performance function from time $t=0$ to time $t=50$. As shown in this figure, the agent positions reach an equilibrium in the end.

Due to the gradient-based nature of our controller, sometimes the obtained solution may be an undesired local optimum where some pair of agents are collocated. In order to avoid these undesired local optima, one can use modified controller functions to prevent collocation. One possible way is to modify the controller as $\dot{p}_i = -\frac{\partial \mathcal{H}}{\partial p_i} + u_i(\|p_i,p_j\|)$, where $u_i(\|p_i,p_j\|)$ is a control term to prevent the collocation (see e.g. \cite{hussein2007effective}).
Such modifications also have much practical significance. For example, the performance of a bi-static radar is very poor when the transmitter is collocated with the receiver. Furthermore, for mobile agents, it is always desirable to keep a safe distance between each pair of agents so as to avoid collision.

\section{Real world applications}
\label{Real world applications}
\subsection{Supermarket problem}
\label{supermarket problem definition}
As an extension of the well-known postoffice problem \cite{clarkson1985probabilistic}, the supermarket problem brings in the idea of more than one supermarket serving one cell which enables competition. When selecting sites for supermarkets, it has to be considered important to ensure enough competition among supermarkets \cite{martin2009review}. In fact, in the Australian Capital Territory, the government issued  a policy saying that supermarkets should be located so that everybody is as near as possible to \textit{at least two} supermarkets.
Although supermarkets are not mobile, the optimization technique is still applicable.
Denote the set of positions of supermarkets as $(p_1, p_2,\cdots,p_n)$ in a region $Q$. For each person at position $q\in Q$, the $f(\cdot,\cdot)$ function is
$$f(\cdot,\cdot)= \max(\|q,p_i\|,\|q,p_j\|)$$
By interpreting $\phi(q)$ as  the population density at each position $q$, one can formulate the following coverage performance function
$$\int_Q \min_{(i,j)\in C} f(\cdot,\cdot) \phi(q) dq$$
where the definition of $C$ is the same with Section \ref{order two section}.
To optimize the supermarkets locations, one should minimize the above performance function.

\subsection{TDOA sensors and bearing-only sensors}
The above \textit{supermarket problem} is not strictly a mobile agents problem. Now we are going to look at some mobile sensing problems using the same distance function with the above.
If one uses bearing only sensors to localize targets, at least two sensors are required to obtain a measurement. Furthermore, for complete coverage, one actually needs three non-collinear bearing-only sensors to avoid the case when the sensors and target are almost collinear.
Similar to bearing-only sensors, TDOA (Time Difference of Arrival) sensors are passive sensors and at least two sensors are required to obtain a branch of a hyperbola on which a target lies. Therefore, to localize a target using TDOA sensors, three sensors at different positions are required. In this case, an order three Voronoi partition may be used to solve the problem.
\footnote{Similar to the bearing-only sensors case, for complete coverage, one actually needs four TDOA sensors to avoid undesired geometry, in which case the use of even higher order Voronoi partition is required. }

In the above example, sensors all have limited sensing range and their performance will degrade as the distance from the target to the sensor increases. If one wants to use sensor arrays to cover a large area, an order 3 coverage control problem is an appropriate tool. When detecting a target, all sensors in a cell need to receive signals from the target reliably, with the performance of monitoring a cell depends on the sensor with furthest distance to the target.  Thus the performance of a group of sensors $p_1, p_2,\cdots,p_n$ monitoring a target at $q$ can be expressed as
$$f(\cdot,\cdot,\cdot)=\max(\|q,p_i\|,\|q,p_j\|,\|q,p_k\|)$$

Suppose there is a target in a two-dimensional region $Q$. Suppose further that the probability of the target appearing at position $q\in Q$ is $\phi(q)$. Then the expected value of this localization performance measure is
$$\int_Q \min_{(i,j,k)\in C_3} f(\cdot,\cdot,\cdot) \phi(q) dq$$
where $C_3=\{i,j,k|~i,j,k\in\{1,\cdots,n\},i < j<k\}$.
Minimizing the above performance function  will give the optimal positions of sensors.

\subsection{Bi-static radar}
One possible application of our order two coverage control strategy involving a general $f(\cdot,\cdot)$ function is the coverage design using bi-static radars \cite{nezlin2007bistatic}. When deploying the a bi-static radar, it is required that the transmitter and receiver are at different locations. We suppose that each agent with position $p_i$ has both transmitter and receiver on board but it can only receive signals from another agent instead of itself. In this case, the probability of detection in one radar pulse is expressed by
\begin{equation}
\label{}
P=\int_{v_t}^\infty \frac{1}{2}I_0(\frac{\delta}{a}\sqrt{\theta})exp(-\frac{\theta-a^2/\delta^2}{2})d\theta
\end{equation}
where the term being integrated is just the probability density of the received signal strength using a common radar receiver model; see \cite{mahafza2013radar}. In particular, $v_t$ is a chosen detection threshold, $a$ is the signal amplitude, $\delta$ is the noise amplitude and $I_0$ is the modified Bessel function of the first kind and zeroth order.
The choice of $v_t$ depends on the acceptable probability of false alarm denoted by $P_{fa}$, i.e. the acceptable probability that the signal magnitude exceeds the threshold value when noise alone is present \cite{mahafza2013radar}.

For a typical bi-static radar, $a^2/\delta^2$ is the signal-to-noise ratio, which is proportional to $K/(R_1^2 R_2^2)$ where $R_1$ is the distance between the transmitter and the target, $R_2$ is the distance between the receiver and the target and $K$ depends on the radar power, antenna gain etc. Note $R_1$ and $R_2$ are equivalent to $\|q,p_i\|$ and $\|q,p_j\|$ in Section \ref{order two section}. It can be shown that
$$\frac{\partial}{\partial R_1}P \leq 0,~\frac{\partial}{\partial R_2}P \leq 0\,\,\,\,\text{and}\,\,\,\, P(R_1,R_2)=P(R_2,R_1)$$

Suppose there are a group of agents to be deployed in a convex area $Q$;  each agent with position $p_i$ has both transmitter and receiver on board but it can only receive signal from another agent instead of itself.
There is a target in the area and the probability of the target appearing at position $q\in Q$ is $\phi(q)$. Note there holds $\int_Q \phi(q)dq=1$. Now the probability of detection of the target at $q$ is $\max_{(i,j)\in C} P(\|q,p_i\|,\|q,p_j\|)$ and the expected value of this probability of detection is
$$E=\int_Q \max_{(i,j)\in C} P(\|q,p_i\|,\|q,p_j\|) \phi(q) dq$$
Now let $f(\cdot,\cdot)=-P(\cdot,\cdot)$, then the performance function $\mathcal{H}$ in Section \ref{order two section} is stated as
$$\mathcal{H}=-E=\int_Q \min_{(i,j)\in C} -P(\|q,p_i\|,\|q,p_j\|) \phi(q) dq$$

The above model is only about the first detection of targets. As for the localization part of this problem, a pair of bi-static radars only gives an eclipse showing the possible positions of a target. If one needs to obtain an exact position of the target, even higher order coverage control strategy is required.

\section{Conclusions}
In this paper, we considered a class of generalized Voronoi coverage control problems by introducing the higher order Voronoi partition concept in the coverage performance functions. This coverage problem is motivated by many real life applications which require more than one sensor to cooperate in monitoring one single cell. We focused on the order 2 Voronoi based coverage problem, and provided detailed analysis on the performance function, the controller design and controller performance, supported by several simulations.  In addition, we provided a number of real world scenarios where our framework can be applied.

In the future, we plan to extend our results to even higher order Voronoi-based coverage control
with more rigorous analysis on the convergence rate. Furthermore, because simulations suggest that away from boundaries of the overall region, the cells are of some standard shape and the area is tiled uniformly, it may be possible to provide some theoretical analysis on the final shape of each cell.
Apart from that, we would also like to use the idea of higher order Voronoi partitions in the research of k-mean clustering, the discretized version of our problem in the field of data analysis.

\begin{appendix}
Here we  recall some basic facts in the problem of differentiation under the integral sign. Suppose $F(x,t)$, the integrand, is a function of both $x$ and $t$; suppose further $D(t)$, the domain of integration, is a function of $t$. Now the derivative of the integral with respect to $t$ at the point $t=t_0$ is
\begin{equation}\label{sperate boundary}
\begin{split}
\frac{d}{dt}\int_{D(t)} F(x,t)dx \Big|_{t=t_0} =
&\int_{D(t_0)} \frac{d}{dt} F(x,t) \Big|_{t=t_0} dx\\
& + \frac{d}{dt}\int_{D(t)}  F(x,t_0)dx \Big|_{t=t_0}\\
\end{split}
\end{equation}
where the second term is the boundary variation. Note the independent variable $x$ could be a vector instead of a scalar in this case. According to \cite{flanders1973differentiation}, the boundary variation equals
$$\int_{\partial D(t)} F(x)v_x\cdot n_x dx$$
where $\partial D(t)$ denotes the boundary of $D(t)$, $v_x$ denotes the direction of moving of a point $x$ on $\partial D(t)$, and $n_x$ denotes the outward unit normal vector of the point $x$ on $\partial D(t)$.
\end{appendix}

\bibliographystyle{IEEEtran}
\bibliography{Coverage}

\end{document}